\documentclass[12pt,a4paper]{article}
\usepackage[T1]{fontenc}
\usepackage[utf8]{inputenc}
\usepackage{booktabs}
\usepackage{amsmath}
\usepackage{graphicx}
\usepackage{hyperref}
\usepackage{xcolor}
\usepackage{siunitx}
\usepackage{lineno}
\usepackage{cite}
\usepackage[export]{adjustbox}
\usepackage{chngcntr}

\begin{document}
\pagenumbering{gobble} %
\title{Daily monitoring of scattered light noise due to microseismic variability at the Virgo interferometer}

\author{Alessandro Longo$^{a}$, Stefano Bianchi$^b$, Guillermo Valdes$^{c,d}$\\ Nicolas Arnaud$^{e,f}$, Wolfango Plastino$^{a,g}$}

\date{\small 
\it{$^a$INFN, Sezione di Roma Tre, Via della Vasca Navale 84, 00146 Rome, Italy.}\\
\it{$^b$ Rome, Italy}\\
\it{$^{c}$ Louisiana State University, Baton Rouge, Louisiana 70803, USA}\\
\it{$^{d}$ Texas A\&M University, College Station, Texas 77843, USA}\\
\it{$^e$ Université Paris-Saclay, CNRS/IN2P3, IJCLab, 91405 Orsay, France}\\
\it{$^f$ European Gravitational Observatory (EGO), I-56021 Cascina, Pisa, Italy}\\
\it{$^g$Department of Mathematics and Physics, Roma Tre University, Via della Vasca Navale, 00146, Rome, Italy}\\}
\maketitle

\begin{abstract}
Data acquired by the Virgo interferometer during the second part of the O3 scientific run, referred to as O3b, were analysed with the aim of characterising the onset and time evolution of scattered light noise in connection with the variability of microseismic noise in the environment surrounding the detector.
The adaptive algorithm used, called pytvfemd, is suitable for the analysis of time series which are both nonlinear and nonstationary. It allowed to obtain the first oscillatory mode of the differential arm motion degree of freedom of the detector during days affected by scattered light noise. The mode's envelope i.e., its instantaneous amplitude, is then correlated with the motion of the West end bench, a known source of scattered light during O3. The relative velocity between the West end test mass and the West end optical bench is used as a predictor of scattered light noise. Higher values of correlation are obtained in periods of higher seismic noise in the microseismic frequency band. This is also confirmed by the signal-to-noise ratio (SNR) of scattered light glitches from GravitySpy for the January-March 2020 period. Obtained results suggest that the adopted methodology is suited for scattered light noise characterisation and monitoring in gravitational wave interferometers. 
\end{abstract}
\clearpage

\section{Introduction}
Virgo is a ground-based gravitational wave (GW) detector located in Cascina, near Pisa (Italy) having a detection frequency band ranging from 10~Hz to a few ~kHz \cite{Acernese_2006,acernese2015advanced}.
Environmental noise characterisation and mitigation are of key importance for detectors such as Virgo. A relevant example in this regard is seismic noise, affecting the detector in the 0.1-200~Hz band \cite{Acernese_2006environmental,Saccorotti_2011seismic,Acernese2004_properties,Effler_2015environmental,Daw:2004qd}. In the 0.1-1~Hz frequency range, the seismic noise induced by sea waves interacting with the shore is referred to as microseism \cite{longuet1950theory,cessaro1994sources}.
Its reduction is fundamental to achieve the design sensitivity of third generation GW interferometers, which aim at enlarging the detection bandwidth at 1~Hz \cite{naticchioni2014microseismic,gorbatikov2008microseismic,bromirski2002near}. For ground-based detectors such as Virgo, the microseismic noise frequency is lower than the resonant frequency of the seismic isolation systems. Hence, an excess of microseismic noise can couple to the detector in the form of scattered light, showing up at the output of the detector i.e., in the differential arm motion (DARM) degree of freedom (DoF) \cite{Accadia_2010scatt,soni2020reducing}.
Scattered light occurs when stray beams bounce off reflective surfaces having a relative motion w.r.t optics of the detector and recombine with the main light beam through phase and amplitude coupling \cite{was2021end}. Its distinctive characteristics are arches, or fringes, appearing in DARMs' spectrograms (see Appendix A). The time at which these arches appear and their frequency $f_{arch}(t)$ is related to the motion $x_{sc}(t)$ of the object responsible for the noise i.e., the culprit, and in particular to its velocity as described by the so-called \emph{predictor} (Hz)
\begin{equation}
f_{arch}(t)=\frac{2N}{\lambda}|v_{sc}(t)|
\label{ffringe}
\end{equation}
where $N$ is the number of round trips occurring before recombination with the main beam, $\lambda$ is the Virgo laser wavelength, and $v_{sc}$ is the velocity at which the detector component is moving. Scattered light is a noise both nonlinear and nonstationary, and adaptive algorithms \cite{Huang_1998,Huang_2014,Wu_2004,Li_2017} are well suited for its characterisation. Equation \ref{ffringe} for the culprit is found to be well correlated with their output \cite{Valdes_2017,Longo_2020sc,GWAS}. Days in which high microseismic noise induced scattered light in the DARM DoF of Virgo were analysed to quantify the contribution of the optical bench located at the end of the West arm, a known source of scattered light during the third scientific run (O3). \emph{Pytvfemd} \cite{stfbnc_2021_4568706}, the Python version of the time varying filter Empirical Mode Decomposition (tvf-EMD) algorithm \cite{Li_2017}, was employed to this end. The analysis focused on the last part of O3 (O3b), the most affected by scattered light, to take advantage of the computing framework at Cascina. The adopted methodology allowed to monitor the onset and time evolution of scattered light noise and link it to the variability of environmental conditions at the detector location.  

\clearpage
\section{Methodology}\label{methodology}
\begin{figure}[!t]
\centering
\includegraphics[scale=0.21]{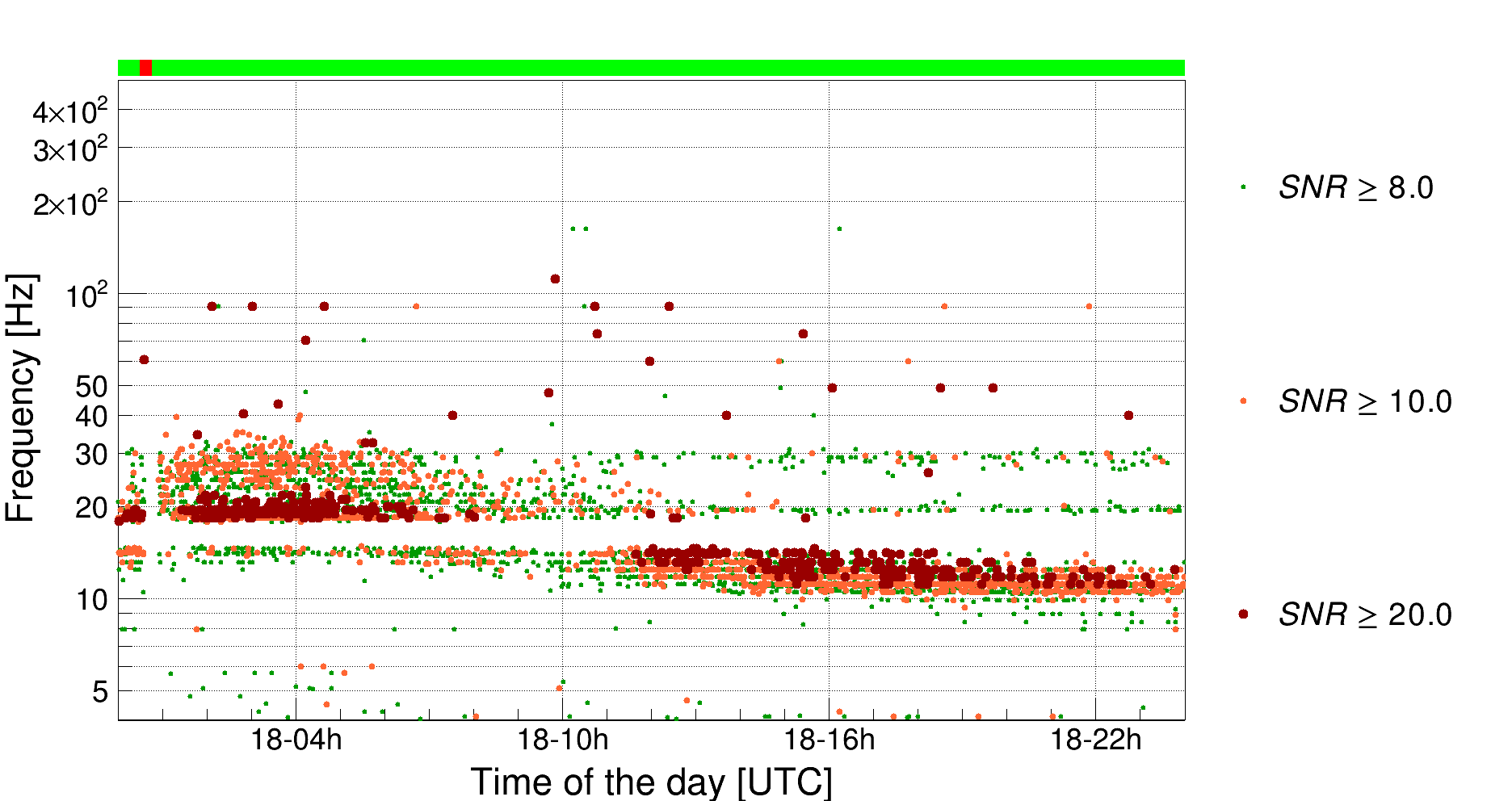}%
\caption{Daily Omicron triggers of DARM for the 18 January 2020. Two clusters of glitches with a high SNR, identified by the larger dots in orange and red, can be observed, one having higher frequency before 11:00UTC and one having lower frequency after 11:00UTC. The difference in frequency of the clusters is consistent with a decrease of the microseismic noise measured at the West end building during the day and reported in Figure \ref{rho}, the scattered light glitches frequency being proportional to the ground motion amplitude.}
\label{omicron}
\end{figure}

Adaptive algorithms allow the extraction of nonlinear, nonstationary modes, referred to as intrinsic mode functions (IMFs), from the data. For this reason, they are suitable for scattered light noise characterisation. When analysing noisy time series with adaptive algorithms such as empirical mode decomposition, mode-mixed IMFs are possibly obtained \cite{Huang_1998}. To mitigate mode mixing, the recently developed tvf-EMD algorithm uses a threshold on the instantaneous bandwidth (IB) of the modes to be extracted to stop the sifting iterations, employing the bandwidth threshold ratio parameter $\xi$. It calculates a local bisecting frequency, used for the spline approximation step which is carried out using B-splines of order $n$ \cite{Unser_1999}. The order $n$ affects the frequency response of the spline approximation (see Figure 1 in \cite{Li_2017}). This bisecting frequency is also realigned (see algorithm 2 in \cite{Li_2017}) to mitigate the effect of noise on the algorithm output. Sifting iterations stop when the oscillatory mode to be extracted is a local narrow band signal, based on a threshold put on the Loughlin IB \cite{Loughlin,Jones_1990}. The relevant parameters of the tvf-EMD algorithm, used for this scattered light data analysis, are $\xi=0.1$ and the B-spline order $n=26$. The Python version of tvf-EMD, referred to as \emph{pytvfemd}, was used. The adopted methodology is based on the one described in \cite{Valdes_2017,Longo_2020sc}, and it was extended taking into account days worth of data. Is hereafter summarised.
\begin{itemize}
\item Periods of scattered light noise in the DARM time series were identified looking at daily Omicron \cite{Robinet_2020omicron} trigger plots such as the one shown in Figure \ref{omicron}. 
\item DARM data were low-pass filtered based on the frequency of the clusters of short transients of excess of power (glitches) with SNR>20, visible in the daily Omicron triggers plots. Then low-passed data were decomposed with \emph{pytvfemd}, obtaining a set of IMFs.
\item The envelope of the first mode (IMF1) i.e., its instantaneous amplitude IA(t), is correlated with Equation \ref{ffringe} computed for the suspended West end bench (SWEB), the most frequent culprit of scattered light in Virgo during O3. The Pearson correlation coefficient is used to this end.
\end{itemize}
The relative motion in the direction of propagation of the beam (z) of SWEB and of the West end test mass (MIRROR) was used, referred to as BENCH-MIRROR.
For each day, intervals of 60 seconds were considered. Hence, the output of the analysis is a time series of 1440 values of correlation $\rho$ for each day. The $\rho$ time series is indicative of the time evolution of scattered light noise due to the West end bench. Its dependence on the microseismic noise, as measured by environmental sensors deployed at the Virgo location, can then be investigated.   

\begin{figure}[!t]
\centering
\includegraphics[scale=0.15,center]{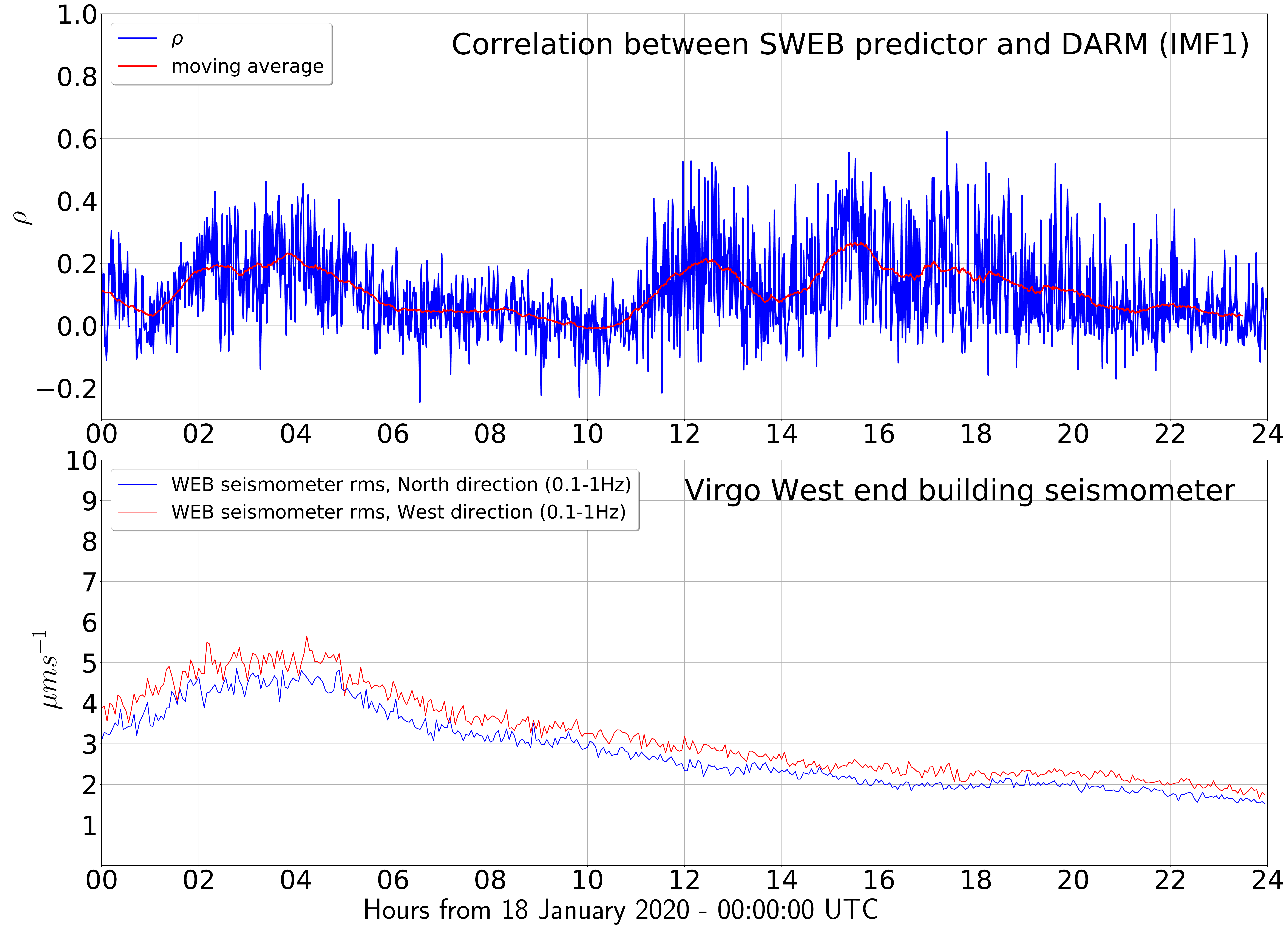}
\caption{Top: correlation values $\rho$ between the predictor $f_{arch}(t)$ for BENCH-MIRROR and the instantaneous amplitude of DARM's first oscillatory mode, as extracted by \emph{pytvfemd}. Bottom: microseismic noise as measured at the Virgo West end building.}
\label{rho}
\end{figure}

\section{Results}\label{results}
This section presents the results of the daily analysis. In total 1440 intervals of 60s of data were analysed for each day. As the winter period is typically characterised by higher sea activity and microseismic noise, the following days of the last part of O3b, were chosen:
\begin{itemize}
\item 18 January 2020
\item 1-2, 9-10, 29 February 2020 
\item 1-2 March 2020  
\end{itemize}
For these days in O3b, a larger number of low frequency triggers were observed in the Omicron trigger data. Figure \ref{omicron} shows the daily Omicron triggers of DARM DoF of Virgo during 18 January 2020. Two distinct clusters of scattered light glitches with a high SNR>20 can be seen. The first cluster has frequencies centered around $20$~Hz, and lasts from 00:00 UTC to approximately 11:00 UTC. The second cluster has frequencies centered around $15$~Hz and it lasts until the end of the day. Figure \ref{rho} shows the correlation $\rho$ between the BENCH-MIRROR predictor $f_{arch}(t)$ and the $IA(t)$ of DARM's IMF1, as extracted by \emph{pytvfemd}. For the 18 of January, a low-pass frequency of $f=20$~Hz was used for the 00:00-11:00 UTC interval while $f=~10$~Hz was used from 11:00UTC to the end of the day. This combination of values was found to give the best results i.e., higher values of $\rho$. It can be seen that the $\rho$ values obtained are consistent with the times of occurrence of high SNR glitches in Figure \ref{omicron}, in the UTC time intervals 01:00 - 07:00, 11:00 - 14:00, and 14:00 - 23:00. The moving average of $\rho$ with a window of 60 points, hence one hour, shown in red highlights this.  
\clearpage

\begin{figure}[!t]
\includegraphics[scale=0.15,center]{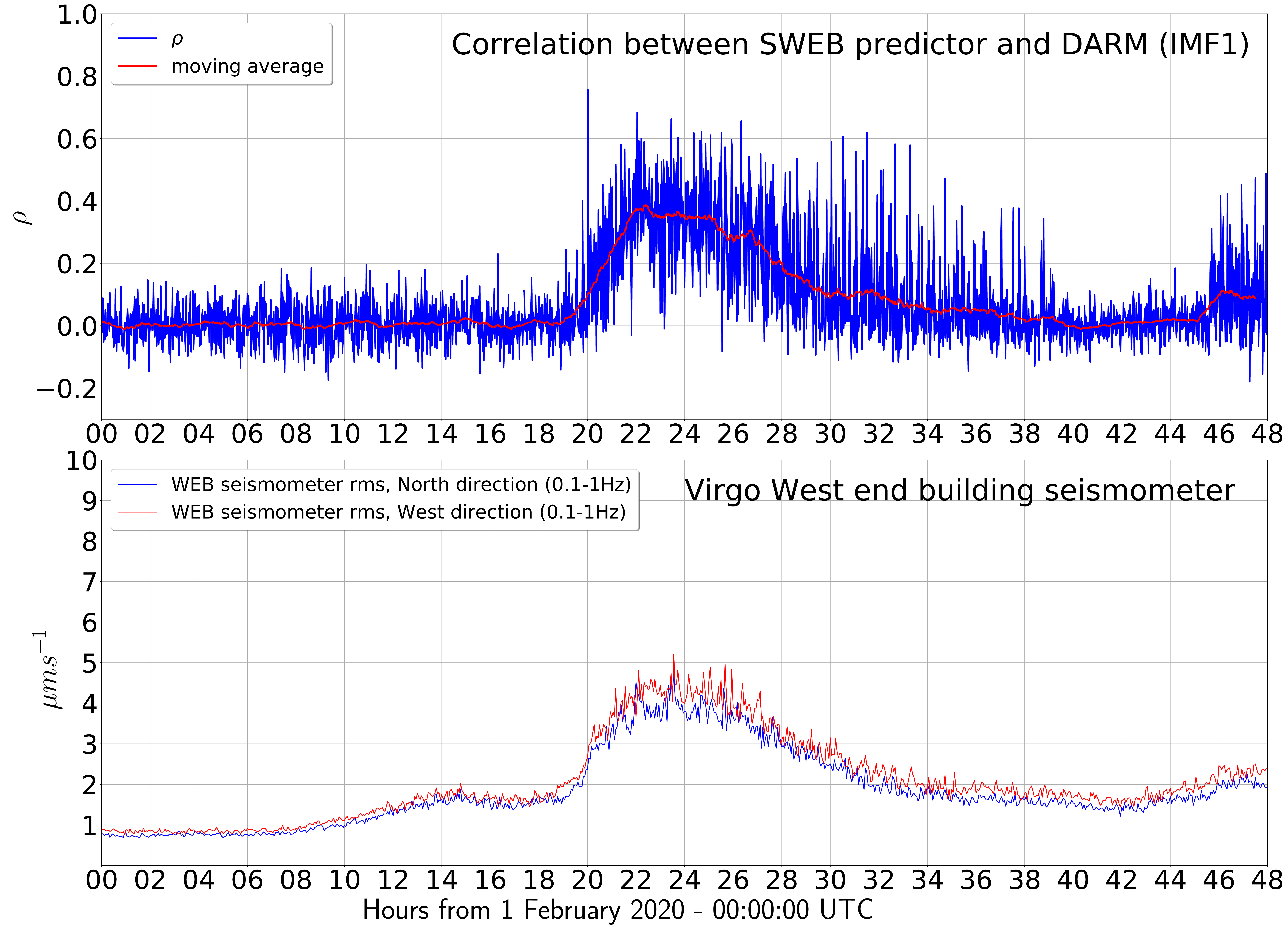}%
\qquad\qquad
\includegraphics[scale=0.15,center]{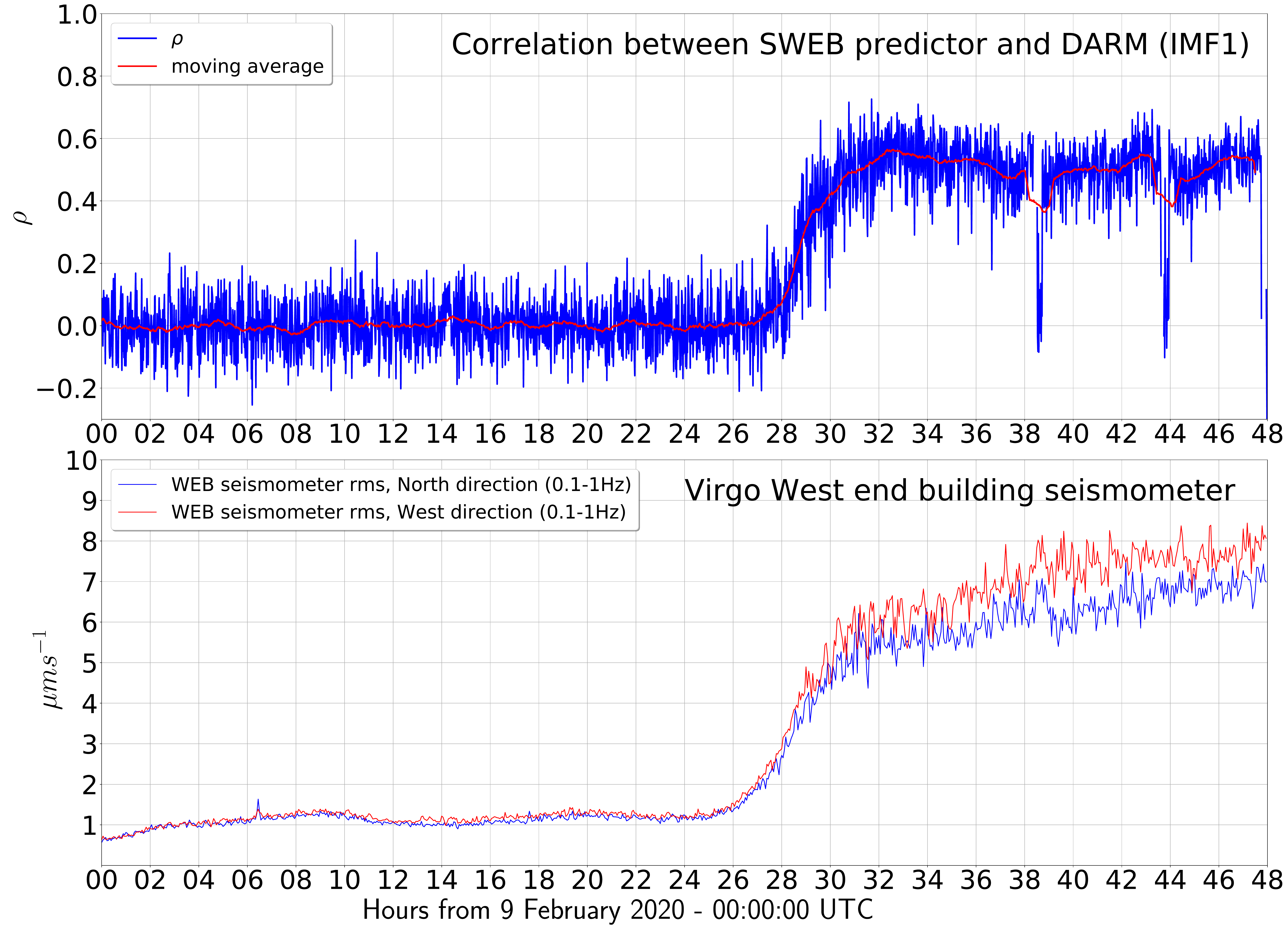}
\caption{Correlation values between the predictor $f_{arch}(t)$ for BENCH-MIRROR and the IA(t) of DARM's IMF1, obtained for the 1-2 February period (top) and the 9-10 February (bottom). Also shown, time evolution of the root means square of the West end building seismometer data in the $0.1-1$~Hz frequency band.}
\label{rho_env}
\end{figure}

\clearpage
\begin{figure}[!t]
\includegraphics[scale=0.15,center]{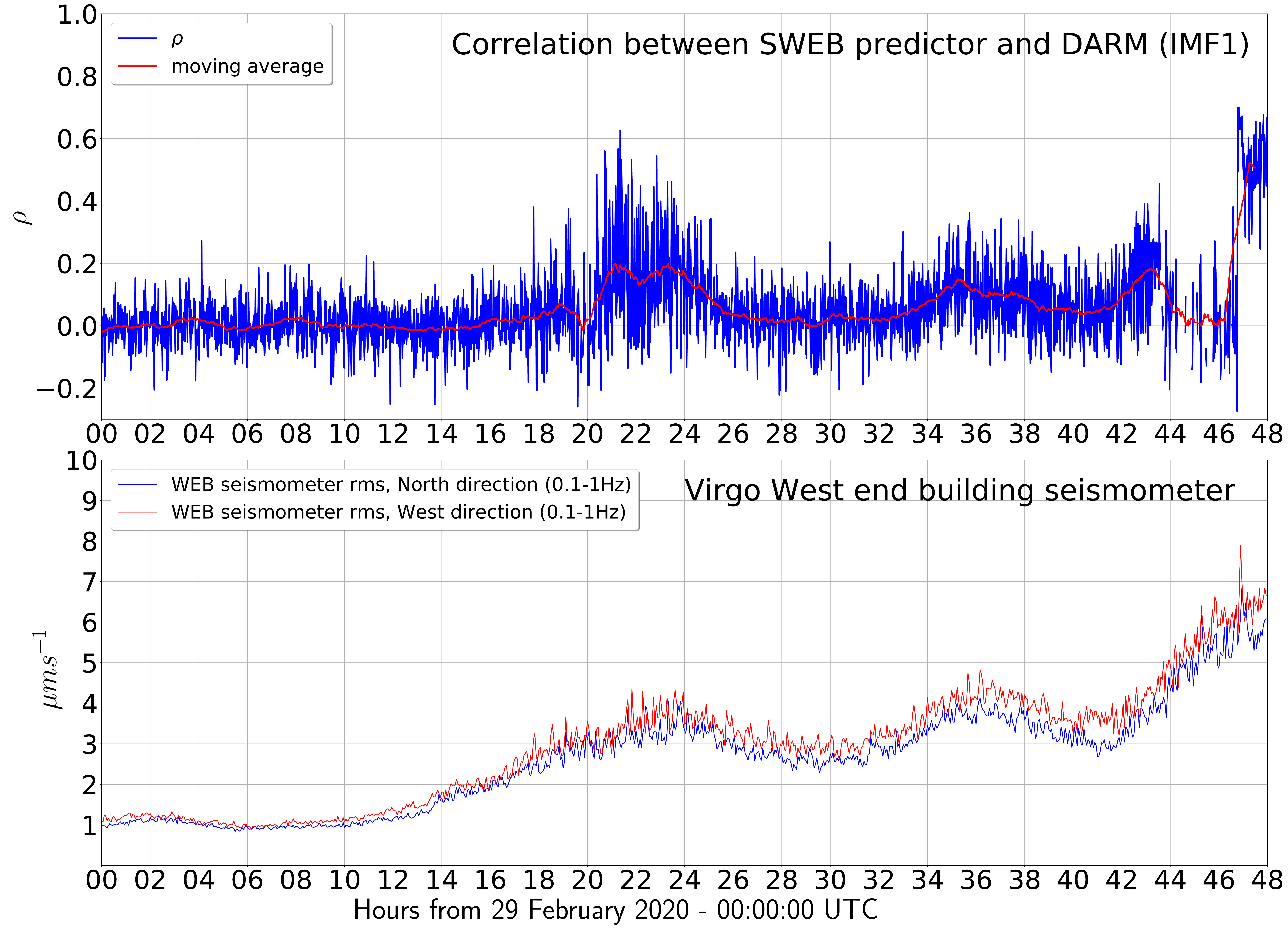}
\includegraphics[scale=0.15,center]{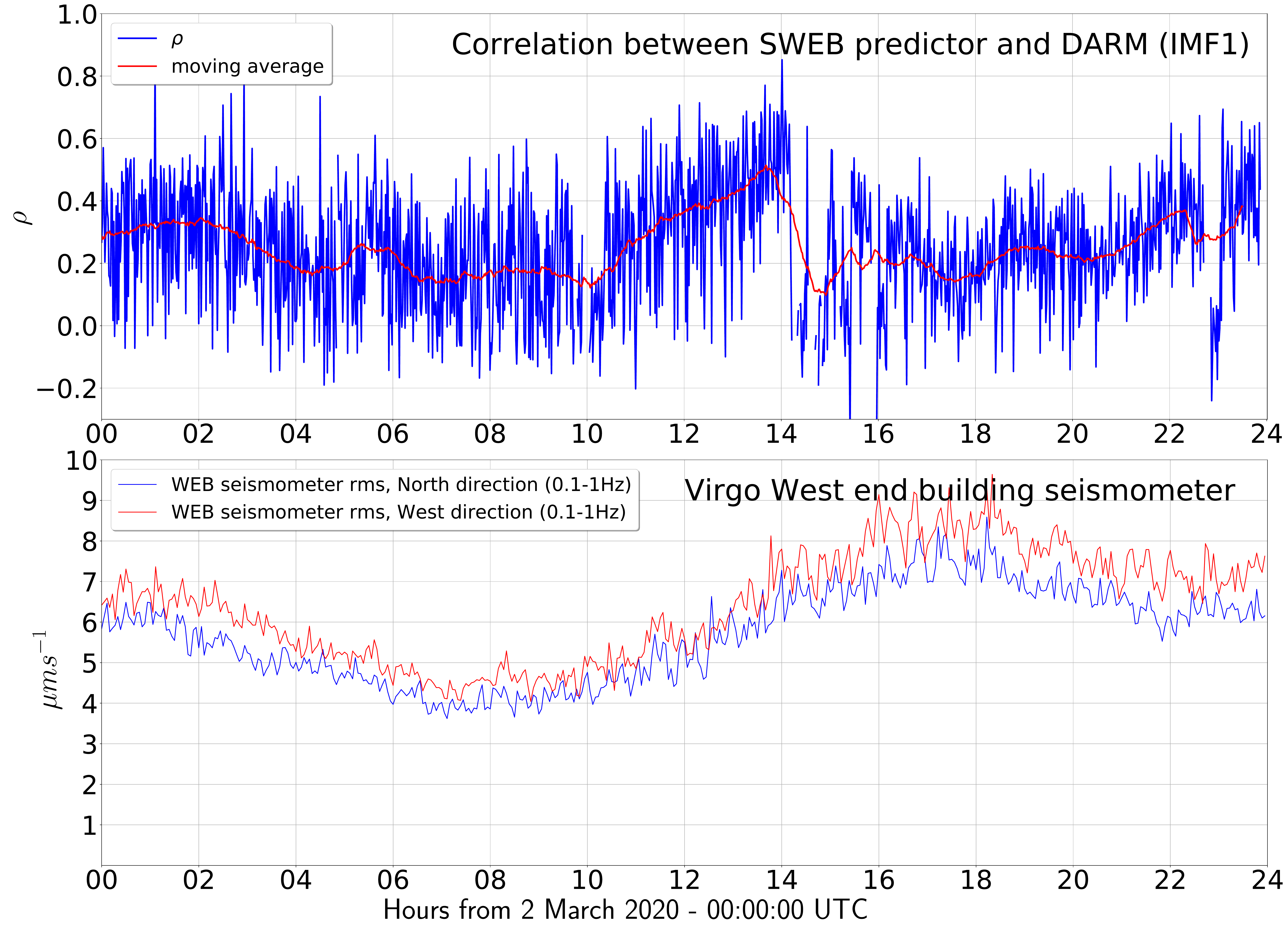}
\caption{Correlation values between the predictor $f_{arch}(t)$ for BENCH-MIRROR and the IA(t) of DARM's IMF1, obtained for the 29 February 1 March period (top) and the 2 March (bottom). Also shown, time evolution of the root means square of the West end building seismometer data in the $0.1-1$~Hz frequency band.}
\label{rho_env2}
\end{figure}

\clearpage

\begin{figure}[!t]
\hspace{-0.5cm}
\includegraphics[scale=0.8,center]{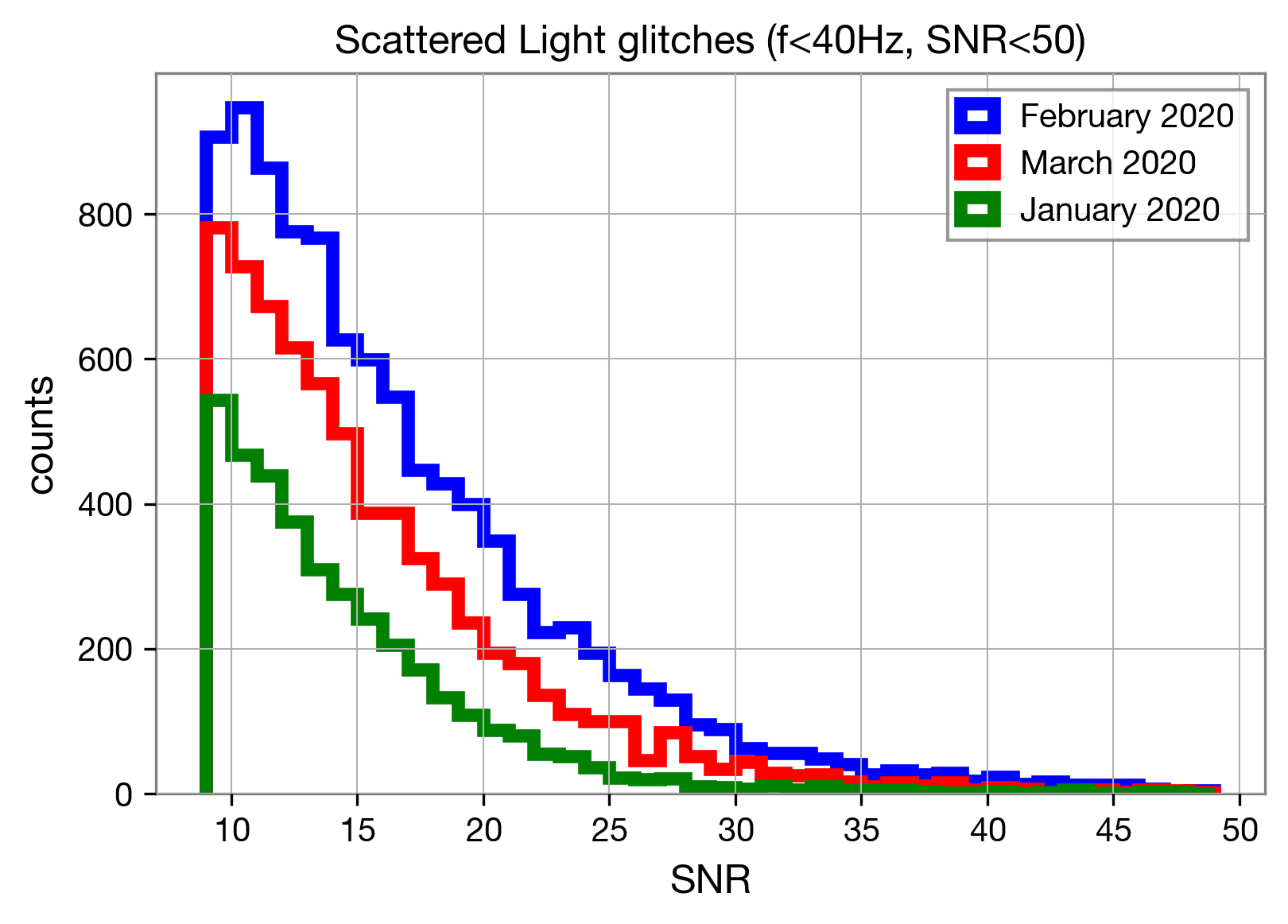}
\caption{Histogram of SNR of Virgo scattered light glitches, with peak frequency $f<40$~Hz and $SNR<50$, during the first three months of 2020. February has the highest number of glitches, followed by March and January.}
\label{glitchesO3}
\end{figure}

Figure \ref{rho}, bottom panel, shows the time evolution of seismic noise in the microseismic frequency region $0.1-1$~Hz at the Virgo West end building on the 18 January. Higher values at the beginning of the day are observed, related to the higher frequency scattered light glitches.
The decrease of microseismic noise is consistent with the observed lower frequency of the second cluster of glitches of Figure \ref{omicron}. Figure \ref{rho_env} shows the $\rho$ values obtained for February 1-2 and February 9-10. A higher correlation, indicative of scattered light noise in the data, is visible in the middle region of the plot. Also shown is the seismometer data, which are highly correlated with the $\rho$ time series of the top panel.
Figure \ref{rho_env2} reports the $\rho$ values obtained for the 29 February - 2 March period, along with seismometer data. Sudden drops in the values of $\rho$, are obtained from data following lock loss of the detector i.e., when not all the optical cavities are resonant, and correspond to periods in which the full lock is not yet achieved. More details on the lock acquisition procedure are in \cite{allocca2020interferometer}. Figure \ref{rho}, \ref{rho_env} and \ref{rho_env2} show that an increase of microseismic noise leads to an increase in scattered light noise witnessed in DARM due to the West end bench i.e., high correlation values between the SWEB predictor and the instantaneous amplitude of DARM's IMF1. Regarding the 2 March, an increase in seismic noise in the $0.03-0.1$~Hz is also measured by the seismometer in the West end building (not shown), coincident with the increase of $\rho$ in the middle part of the plot. To further investigate the relationship between scattered light glitches and microseismicity, Figure \ref{glitchesO3} shows histograms of the signal-to-noise ratio (SNR) of Virgo glitches classified as scattered light by GravitySpy \cite{zevin2017gravity} with $SNR<50$ during January, February and March 2020. Scattered light glitches with peak frequency $f<40$~Hz and for which the duration is shorter than their separation in time were considered. It can be seen that February has the highest number of glitches, followed by March and January. Changes in the amounts of scattered light glitches are possibly due to different levels of microseismicity during the considered months. 

\begin{figure}[!t]
\hspace{-0.5cm}
\includegraphics[scale=0.08,center]{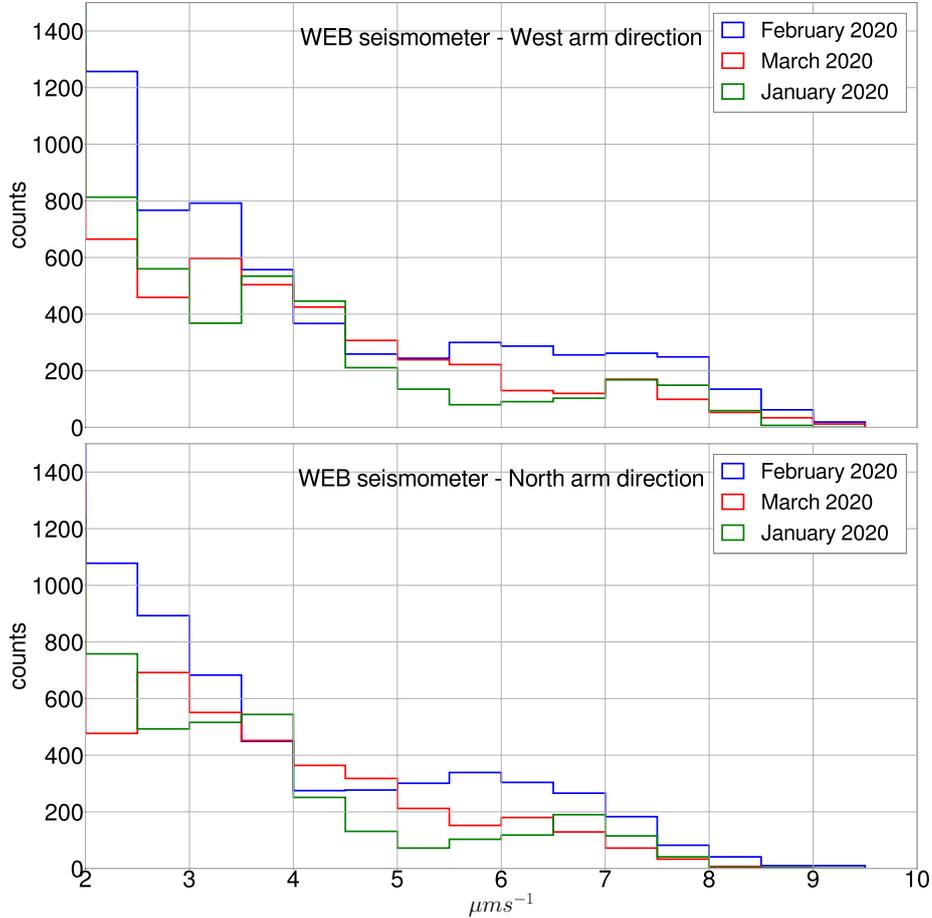}
\caption{Histogram of Virgo West end building seismometer data in the microseismic frequency region, horizontal components. Only velocities greater than $v=2$\si{\mu ms^{-1}} are shown. Velocities greater than $v=5$\si{\mu ms^{-1}} are more frequent in February, wrt March and January 2020.}
\label{seism_month}
\end{figure}

To verify this, Figure \ref{seism_month} reports histograms of the West end building seismometer data, horizontal components, in the microseismic band i.e., $0.1-1$~Hz, for the same period. Velocities greater than $v=5$\si{\mu ms^{-1}} are more frequent in February followed by March and January 2020, as expected.

\section{Conclusions}\label{conclusions}
The \emph{pytvfemd} adaptive algorithm was applied to the DARM DoF of the Virgo detector to extract the oscillatory mode related to scattered light noise, from which its instantaneous amplitude IA(t) can be obtained. The IA(t) time series is known to correlate with the predictor computed from the velocity of the optic generating scattered light. Daily analysis were carried out, using the relative motion of the West end bench wrt the West end mirror to obtain the predictor from Equation \ref{ffringe}. The analysis obtains a total of 1440 values of Pearson correlation coefficients $\rho$ each day. The time evolution of $\rho$ values is consistent with the clusters of scattered light glitches in the Omicron triggers plot of 18 January, shown in Figure \ref{omicron}. Omicron triggers for the other days considered are in Appendix B. Furthermore, the obtained values of $\rho$ are found to be correlated with seismometer data in the microseismic frequency band. The data considered are from the winter period, during which higher microseismic noise is expected. This excess of low frequency noise can couple to the detector in the form of scattered light as it possibly affects the suspended optical benches. Obtained results indicate that the adopted methodology, based on the \emph{pytvfemd} adaptive algorithm, is suitable to monitor the onset and evolution of scattered light noise, related to environmental noise variability in GW interferometers. As the source of scattered light during O3 was known to be the West end bench, this noise was mainly removed during the online noise subtraction using the relevant photodiode as a witness channel \cite{CALIB}. This allowed to remove the scattered light noise so that it would not significantly affect the gravitational wave strain data. The issue concerning the West end bench control was identified and cured after O3, and its residual motion is expected to be similar to the one of the other terminal bench, located at the end of the North Arm of the detector. To make the adopted methodology automated, the expected peak frequency of the scattered light glitches could be used as low-pass frequency cutoff. One daily analysis takes approximately 1.5 hours on Condor \cite{condor}. This allows to foresee a daily monitoring of possible scattered light culprits during the pre-O4 noise hunting phase and then during the run itself. 
\section*{Acknowledgements}
The authors gratefully acknowledge the Italian Istituto Nazionale di Fisica Nucleare (INFN),  
the French Centre National de la Recherche Scientifique (CNRS) and
the Netherlands Organization for Scientific Research (NWO), 
for the construction and operation of the Virgo detector
and the creation and support of the EGO consortium.
The authors also gratefully acknowledge research support from these agencies as well as by 
the Spanish  Agencia Estatal de Investigaci\'on, 
the Consellera d'Innovaci\'o, Universitats, Ci\`encia i Societat Digital de la Generalitat Valenciana and
the CERCA Programme Generalitat de Catalunya, Spain,
the National Science Centre of Poland and the European Union‚ European Regional Development Fund; Foundation for Polish Science (FNP),
the Hungarian Scientific Research Fund (OTKA),
the French Lyon Institute of Origins (LIO),
the Belgian Fonds de la Recherche Scientifique (FRS-FNRS), 
Actions de Recherche Concertées (ARC) and
Fonds Wetenschappelijk Onderzoek‚ Vlaanderen (FWO), Belgium,
the European Commission.
The authors gratefully acknowledge the support of the NSF, STFC, INFN, CNRS and Nikhef for provision of computational resources.
The pytvfemd algorithm was developed in the framework of the Virgo collaboration. The \href{https://pypi.org/project/gwdama/}{\emph{gwdama}} software was used for data access. A. Longo's work is carried out in the framework of the INFN project FIGARO (Fostering Italian Leadership in the Field of Gravitational Wave Astrophysics) and of INFN Virgo Roma Tre. G. Valdes thanks Consejo Nacional de Ciencia y Tecnolog\'ia -  CONACyT of M\'exico for the support provided through the assistantship \textit{Estancias Posdoctorales en el Extranjero}. The authors are grateful to Giuseppe di Biase and the EGO computing department for advice and support. The authors are also grateful to Florent Robinet for providing help with the Omicron software.
\clearpage
\appendix
\counterwithin{figure}{section}

\section*{Appendix A: Scattered light arches in DARM spectrogram}
The distinctive feature of scattered light noise in the data is the presence of arches appearing in their spectrogram. Equation \ref{ffringe} is indicative of the frequency and time of occurrence of such arches. Hence, if the source of scattered light has been correctly identified, the predictor overlaps with the arches in the spectrogram. Figure \ref{counterproof} shows the spectrogram of DARM and the predictor for the West end bench for the 1 February 2020, as obtained for one minute of data. The starting time is 20:01:00 UTC. It can be seen that the predictor overlaps with the arches, confirming that the West end bench is the culprit of the observed scattered light noise.

\renewcommand{\thefigure}{A\arabic{figure}}
\setcounter{figure}{0}

\begin{figure}
\includegraphics[scale=0.8,center]{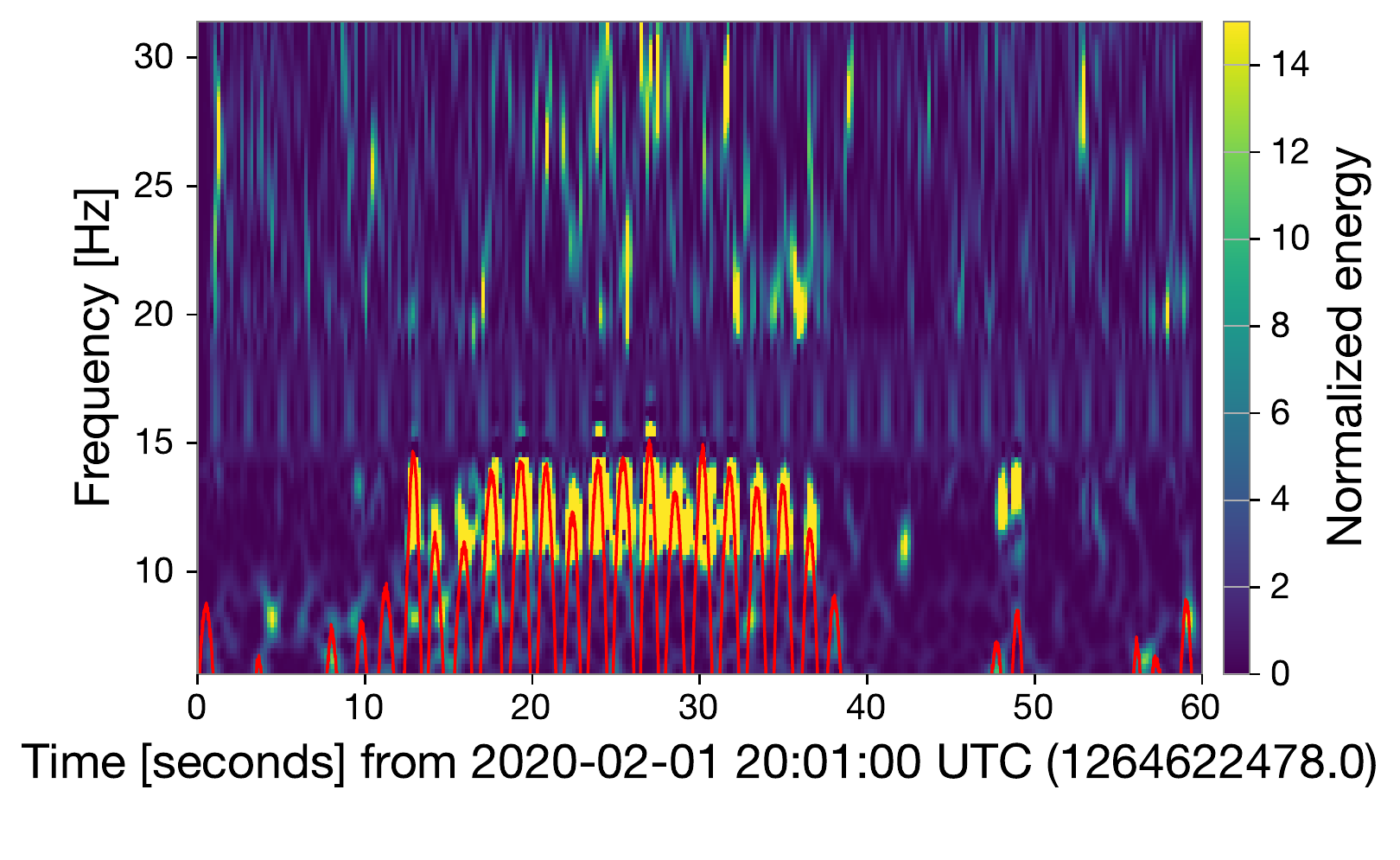}
\caption{Shown in red is the predictor $f_{arch}(t)$ computed for the West End Bench and overlapped on the spectrogram of the DARM DoF of the detector. The scattered light arches are visible in yellow, having frequency up to $15$~Hz.}
\label{counterproof}
\end{figure}

\clearpage

\section*{Appendix B: Daily Omicron plots}
Shown are the Omicron daily triggers of Virgo DARM DoF for the days considered in this study. The onset and time evolution of low frequency glitches is visible. The colour scale indicate the signal-to-noise ratio (SNR) of the glitches. The green and red bar on top indicates periods in which the detector is locked or unlocked, respectively. 

\renewcommand{\thefigure}{B\arabic{figure}}
\setcounter{figure}{0}

\begin{figure}[!b]
\hspace{-0.5cm}
\includegraphics[scale=0.21,center]{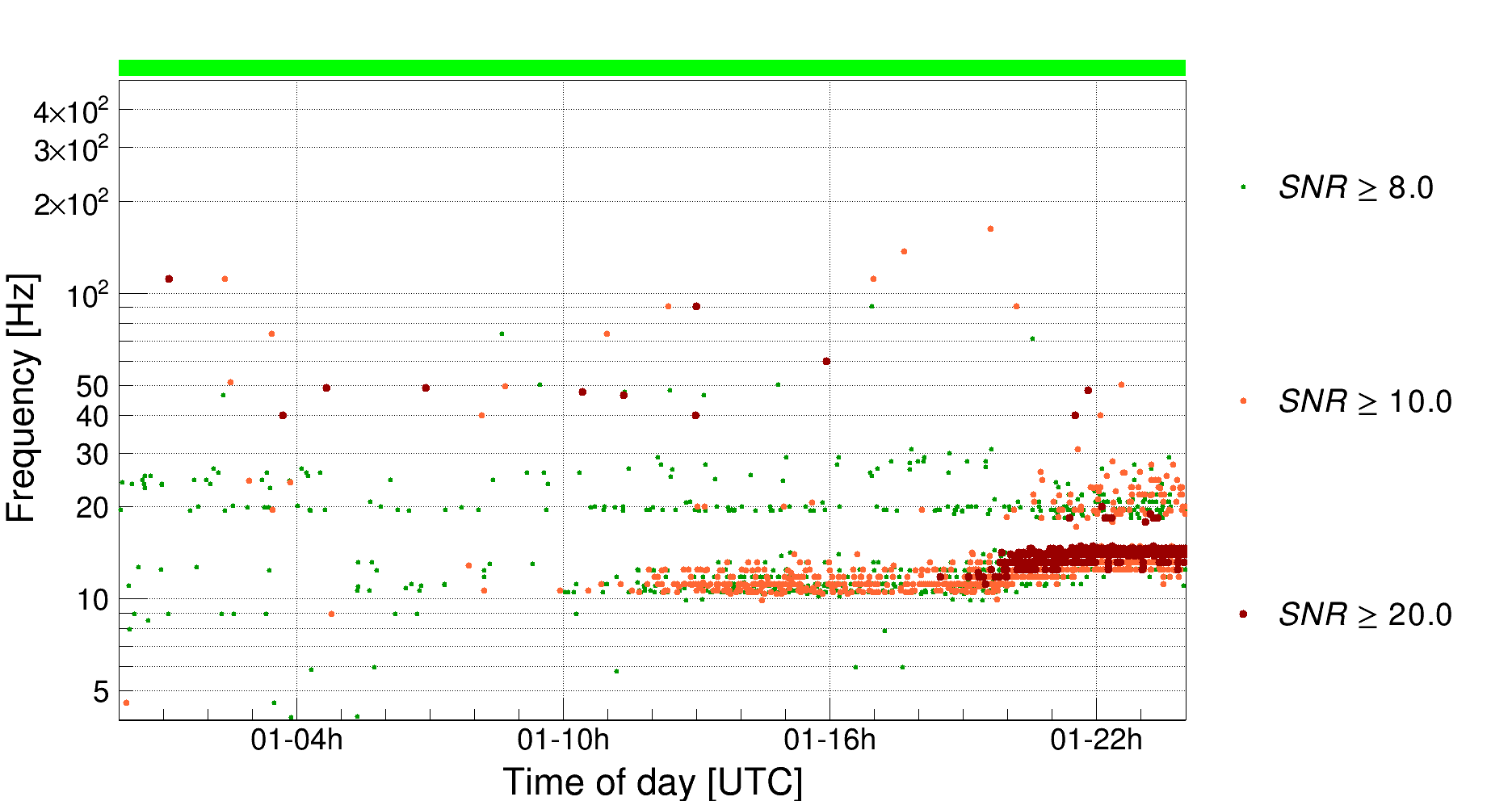}%
\qquad\qquad
\includegraphics[scale=0.21,center]{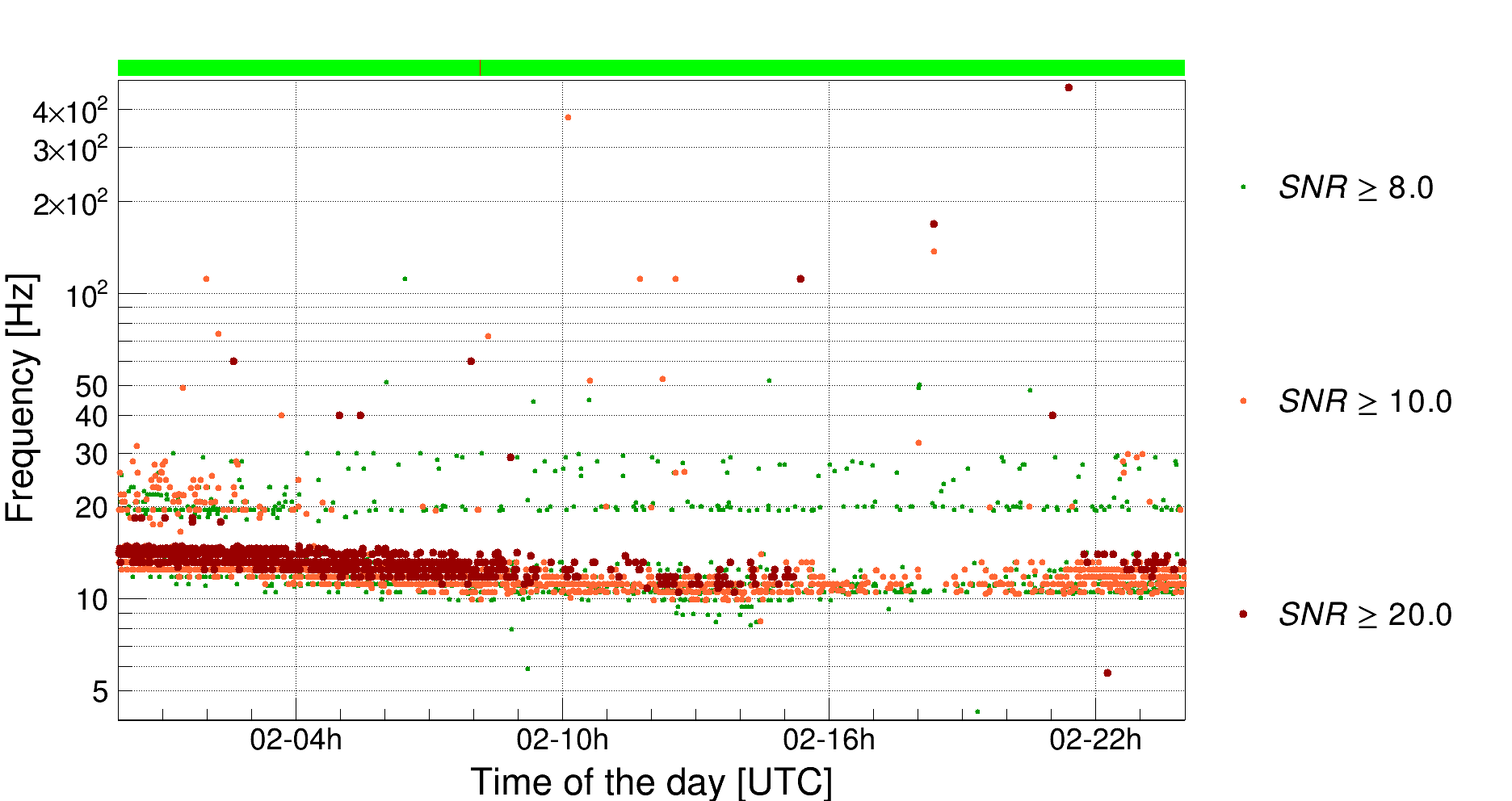}
\caption{The daily Omicron plots for the 1 of February 2020 (top) and for the 2 of February 2020 (bottom) are shown.}
\label{omicron_1_2feb}
\end{figure}

\clearpage

\begin{figure}[!t]
\hspace{-0.5cm}
\includegraphics[scale=0.2,center]{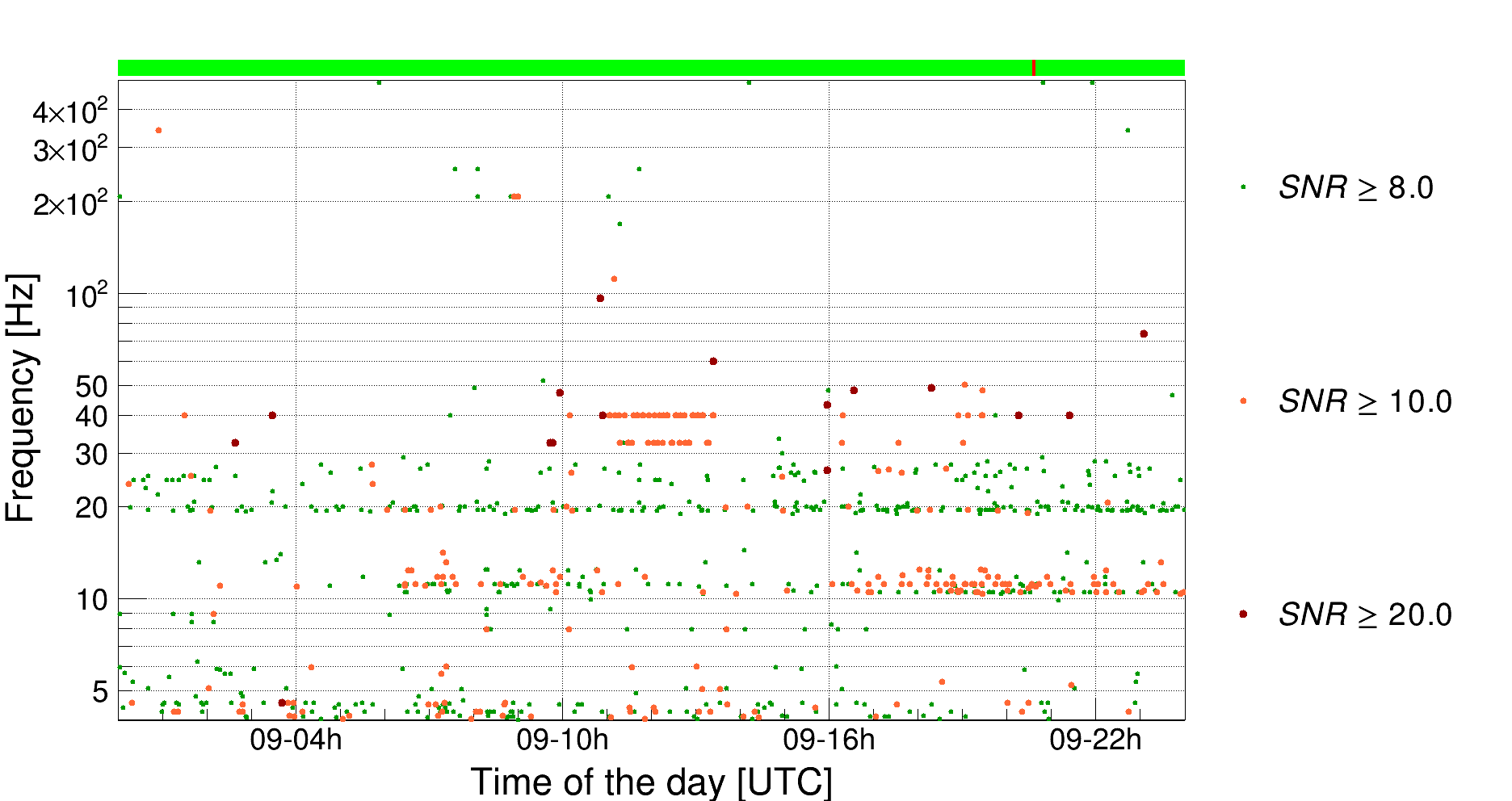}%
\qquad\qquad
\includegraphics[scale=0.2,center]{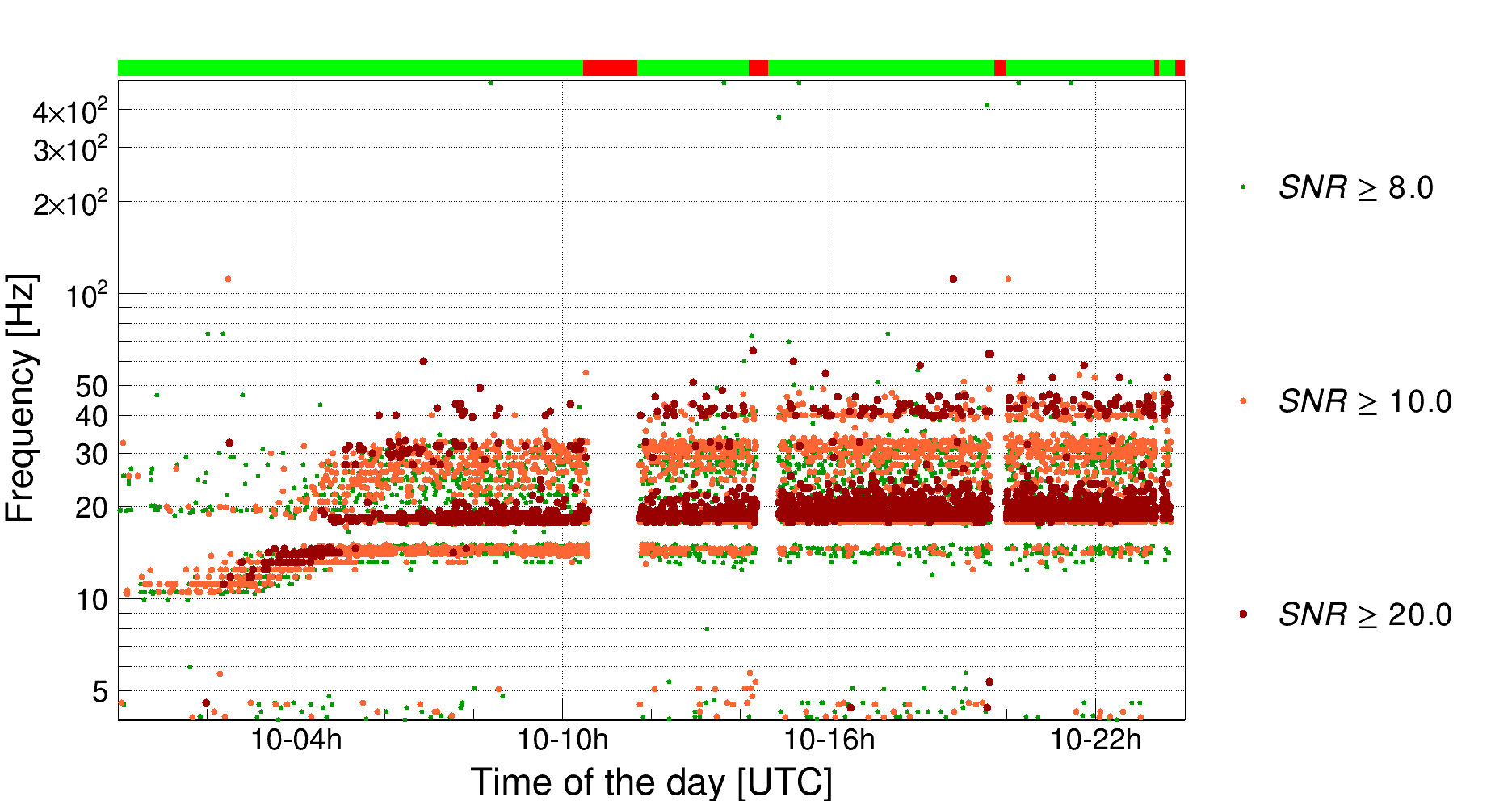}
\qquad\qquad
\includegraphics[scale=0.2,center]{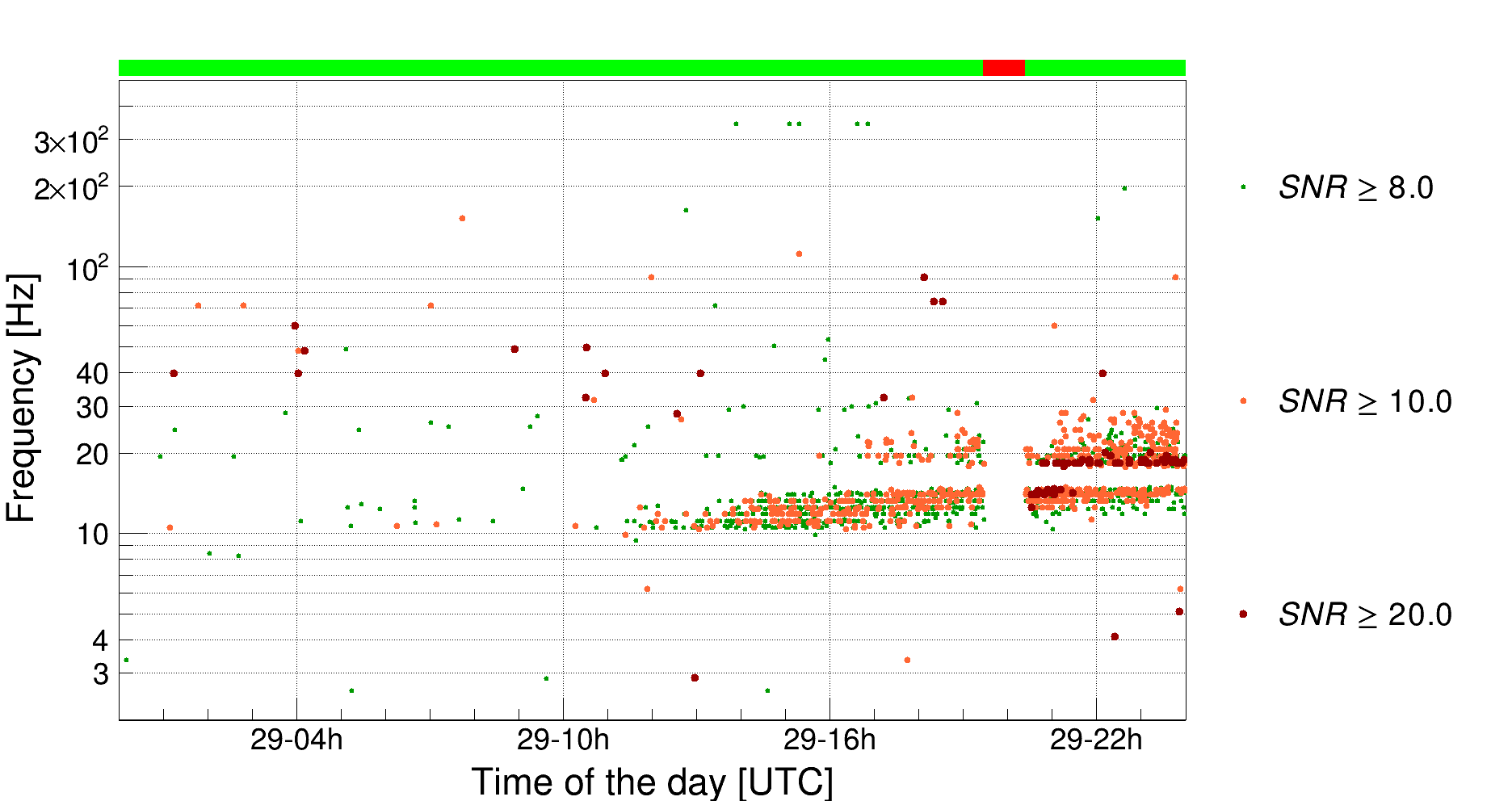}
\caption{From top to bottom, reported here are the daily Omicron plots for the 9, 10 and 29 February 2020.}
\label{omicron_9_10feb}
\end{figure}

\clearpage

\begin{figure}[!t]
\hspace{-0.5cm}
\includegraphics[scale=0.2,center]{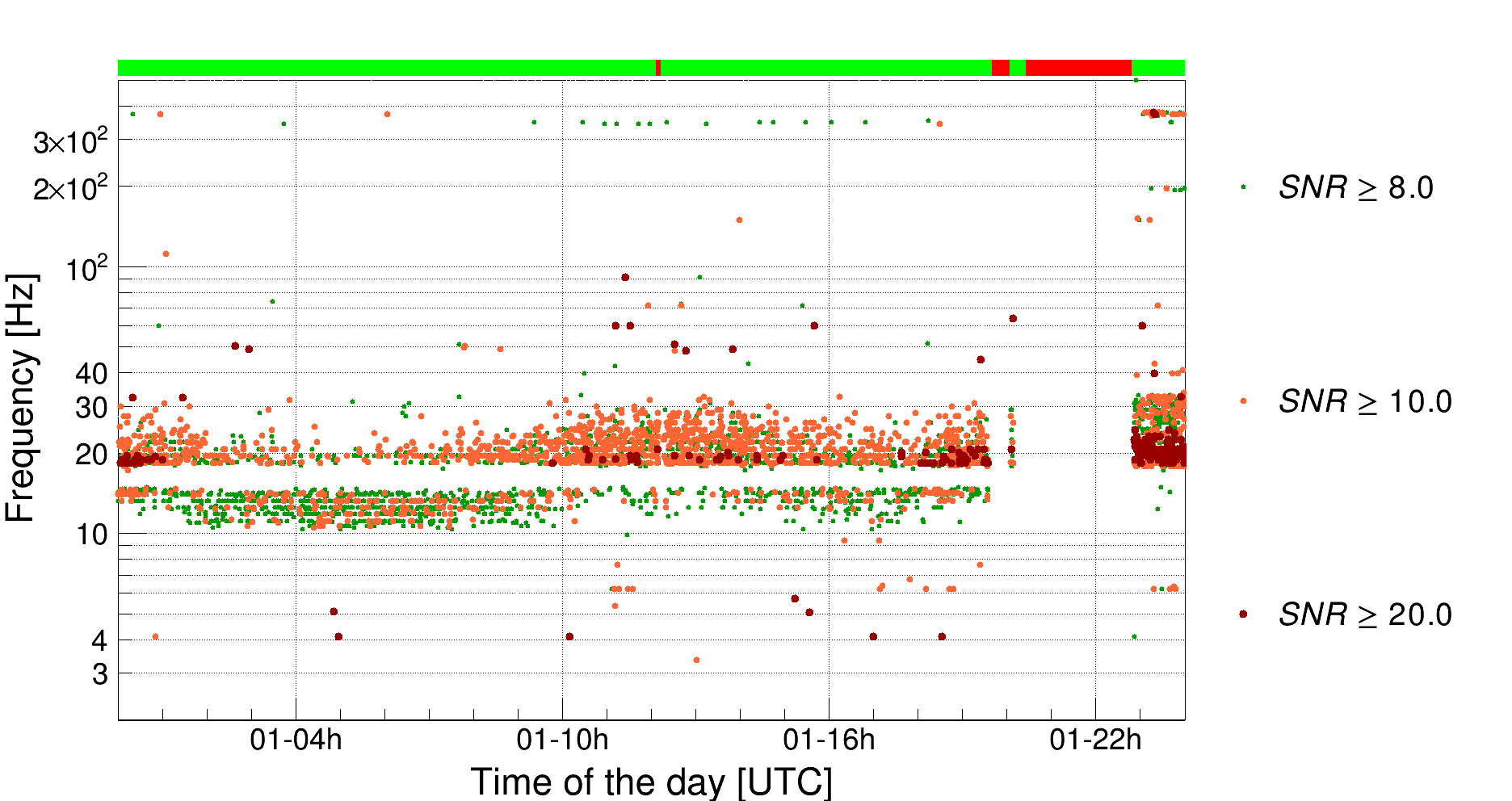}%
\qquad\qquad
\includegraphics[scale=0.2,center]{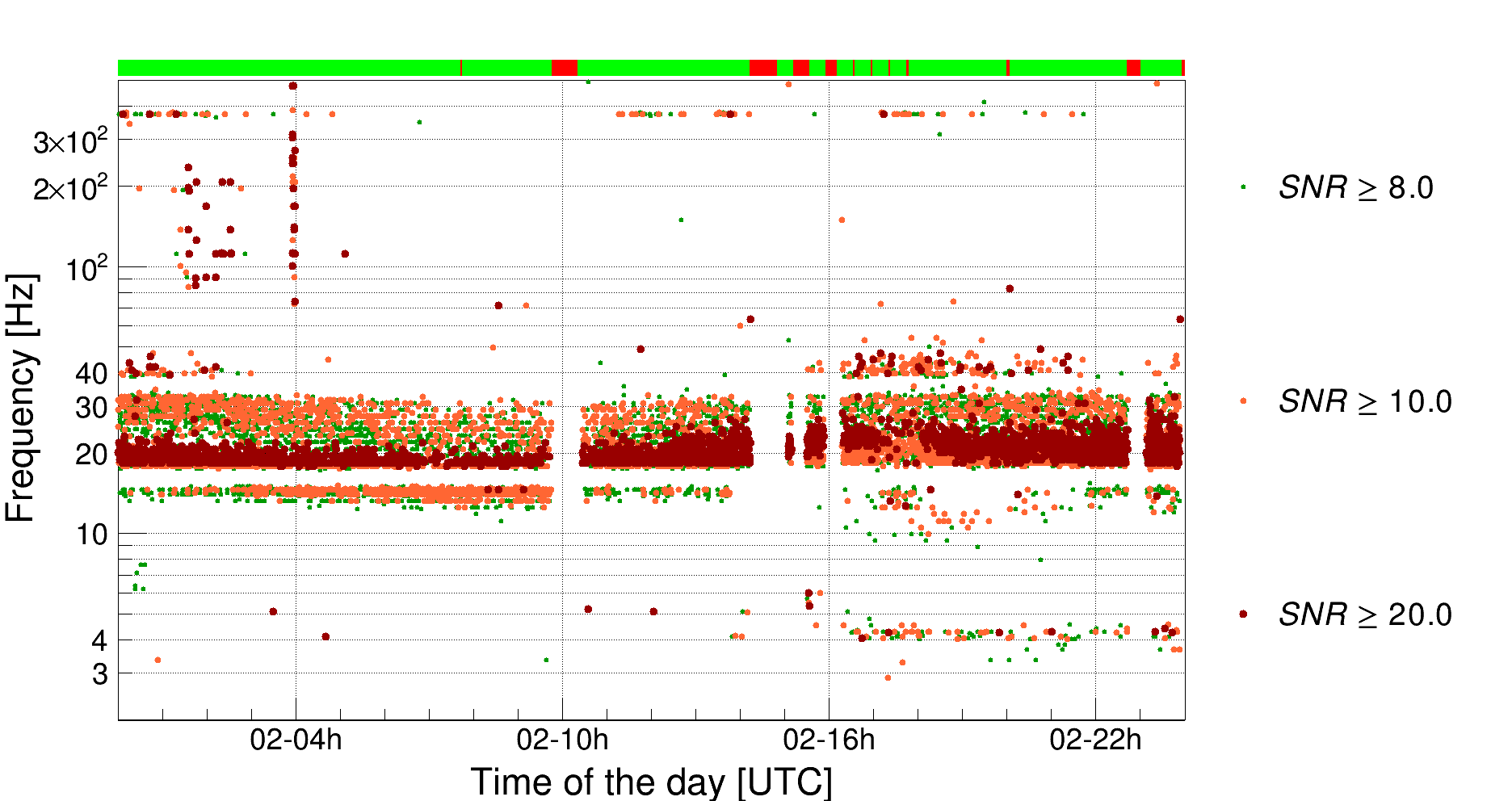}
\caption{Daily Omicron plots for the 1 (top) and 2 (bottom) of March 2020 are shown.}
\label{omicron_1_2mar}
\end{figure}

\bibliography{bibl}
\bibliographystyle{ieeetr}

\end{document}